\documentclass{article}

\usepackage{iclr2023_conference,times}
\iclrfinalcopy % Uncomment for camera-ready version, but NOT for submission.
\usepackage[utf8]{inputenc} % allow utf-8 input
\usepackage[T1]{fontenc}    % use 8-bit T1 fonts
\usepackage{hyperref}       % hyperlinks
\usepackage{url}            % simple URL typesetting
\usepackage{booktabs}       % professional-quality tables
\usepackage{amsfonts}       % blackboard math symbols
\usepackage{nicefrac}       % compact symbols for 1/2, etc.
\usepackage{microtype}      % microtypography
\usepackage{xcolor}         % colors
\usepackage{graphicx}
\usepackage{subfigure}

\definecolor{pastel_blue}{rgb}{0.86, 0.926, 0.984}

\usepackage{adjustbox}
\usepackage[frozencache,cachedir=.]{minted} %to use cached files
\usepackage{xcolor}
\usepackage{array}
\usepackage{multicol}
\usepackage{siunitx}
\usepackage{xspace}
\usepackage{ulem}
\usepackage[T1]{fontenc}
\usepackage{listings} %advanced verbatim
\setminted{fontsize=\scriptsize}
\usepackage{textcomp}

\NewDocumentCommand{\ShowInline}{v}{%
#1%
}

\definecolor{bg}{HTML}{282828} % from https://GitHub.com/kevinsawicki/monokai
\usepackage{pmboxdraw}
\usepackage{footmisc}

\newcommand*{\score}[1]{\num[round-mode=places,round-precision=1]{#1}}
\newcommand*{\scorethree}[1]{\num[round-mode=places,round-precision=3]{#1}}

\newcommand*{\compacc}[1]{\num[round-mode=places,round-precision=1]{#1}\%}

\newcommand \samplingtemp {1.3\xspace}

\usepackage{etoolbox}

\newcommand{\github}{GitHub\xspace}
\newcommand{\codenet}{CodeNet\xspace}

\newcommand{\infer}{Infer\xspace}

\begin{document}
\title{Leveraging Static Analysis for Bug Repair}
\makeatletter
\newcommand{\printfnsymbol}[1]{%
  \textsuperscript{\@fnsymbol{#1}}%
}
\makeatother

\author{\textbf{Ruba Mutasim} \quad
\textbf{Gabriel Synnaeve} \quad
\textbf{David Pichardie} \quad
\textbf{Baptiste Rozière}\\ %Author order TBD (Ruba is first of course)
\quad\hspace{4.5cm}\textbf{Meta AI}\\
\quad\hspace{4.15cm}\textbf{broz@meta.com}
}

\begin{center}
\maketitle
\end{center}

\begin{abstract}
We propose a method combining machine learning with a static analysis tool (i.e.~\infer) to automatically repair source code. 
Machine Learning methods perform well for producing idiomatic source code. 
However, their output is sometimes difficult to trust as language models can output incorrect code with high confidence. 
% ML models are good at producing idiomatic code but not trustable. 
Static analysis tools are trustable, but also less flexible and produce non-idiomatic code. 
In this paper, we propose to fix resource leak bugs in IR space, and to use a sequence-to-sequence model to propose fix in source code space. 
We also study several decoding strategies, and use \infer to filter the output of the model.
On a dataset of \codenet submissions with potential resource leak bugs, our method is able to find a function with the same semantics that does not raise a warning with around 97\% precision and 66\% recall. 
% We can verify these fixes for X functions, and guarantee that the output of our method is correct assuming that \infer is.
% Potential to remove a huge barrier for the use of ML to debug code. 
% Some other results. 
% We rely on the trustable outputs of \infer to find buggy functions, and generate automatic fixes in the IR space. 
% Then, we train a neural network to inverse these IRs and generate fixed source code. 
% The functions we obtained can be compiled again, and compared to the repaired IR which we know is correct. 
\end{abstract}

\section{Introduction}
\label{sec:introduction}

Detecting and repairing bugs in source code necessitates strong reasoning skills over code structures. Current models struggle to capture such reasoning in the absence of a sufficiently large and reliable dataset of bug fixes. Several methods rely on mining \github to extract bugs and their corresponding fixes using some forms of heuristics \citep{tufano2019empirical,chen2019sequencer,gupta2017deepfix}, which can often introduce noise into the extracted dataset.
Other works rely on synthetically created bugs~\citep{hellendoorn2020global,allamanis2021self,pradel2018deepbugs,yasunaga2020graph}. While they are able to generate a large amount of parallel data, their examples do not always match real bugs found in real code. %there is no guarantee that the bugs are representative of real bugs found in real code. 
There are also some fundamental challenges associated with using machine learning models for bug repair. These challenges stem from the fact that these models can make subtle or unexpected errors, and it is often difficult to trust the output of a machine learning model without manual checks. 
Static analysis tools, on the other hand, produce reliable results, but are frequently limited in scope and rarely capable of proposing idiomatic fixes to the source code.
% David: paragraph about Infer

Infer is an open-source static analysis tool developed at Meta and used in industrial settings. It targets mobile apps written in Java/Kotlin or Objective C, and also C++ code used in website backends. It already reported 100,000 issues~\citep{Infer:DistefanoFLO19}. 
At the moment, bugs revealed by \infer are fixed manually by developers. Tools proposing automatic code repair would help improve the bug-fixing rate. 
This often requested feature is difficult to add to static analysis tools like Infer, because they do not directly analyze source code, but rather a low level representation (similar to LLVM IR~\citep{LLVM:CGO04}). In some situations, proposing a repair at IR level is feasible, but still useless from the programmer point of view, as there are no tools to decompile IR to source code.

%. There does not exist decompilation transformation from IR to source code.   
%Many static analysis tools, such as \infer, first compile the code into an Intermediate Representation (IR) in which the analysis is performed. 
%Bugs are often easier to find and fix in IR space than in the source code. 

 % Motivated by the challenges associated with end-to-end learned program repair mentioned above, and in order to leverage the reliability of bug detection and correction in static analysis tools, 
 In this paper, we propose using machine learning in conjunction with infer (static analysis tool) to fix bugs and repair code. This approach can be used to solve bugs in real code. 
 It leverages neural machine translation to generate fixed source code, while deferring the reasoning about bug detection and correction to Infer, which is much more reliable in the absence of high-quality code repair data.
More precisely, we train a model to decompile \infer IRs to code, and use it to retrieve repaired source code from repaired IRs. 

Our main contributions are: 
\begin{itemize}
\item We generate a parallel dataset of Java code and \infer IRs, and train a model to decompile \infer IRs. 
\item We propose a reliable method to automatically fix resource allocation bugs in the \infer IR.
% \item We train a model to decompile \infer IRs, and show that we obtain source code with the same semantics in most cases. 
\item We combine the new auto-fix tool and the decompiler to repair resource allocation bugs and recover fixed java source code. We use our new method to fix more than 66\% of the \infer warnings in our dataset. The proposed fixes still pass the unit tests for 96.9\% of the submissions. 
% \item We also propose to re-compute the IR of our proposed fixes and to compare them with the fixed IR. We report a mean normalized IR edit distance of X over all of our fixes. \br{will we be able to get that?} 
\end{itemize}

% \begin{figure}[ht]
% \vskip 0.2in
% \begin{center}
% Example of what our model can do showing the pros of our method.
% \caption{\small\textbf{Example of bug solved by our method.}}
% \label{teaser_example}
% \end{center}
% \vskip -0.2in
% \end{figure}

\section{Method}
\label{sec:method}
This section describes our method for automatically fixing resource leak bugs flagged by \infer in Java. 
We describe our dataset, modeling choice, and how the fixes are done automatically in the IR. 
Our modelization choices for IR inversion are largely inspirated by~\citet{szafraniec2022code}, who train a model decompiling LLVM IRs for code translation. 
% \br{Explain the IR inversion and RL methods. Make schemas. }

% \makebox[\textwidth][c]{\includegraphics[width=1.2\textwidth]{transcompilation4.pdf}}%

\begin{figure}[ht]
\vskip 0.2in
\begin{center}
\makebox[\textwidth][c]
{\includegraphics[width=1.2\textwidth]{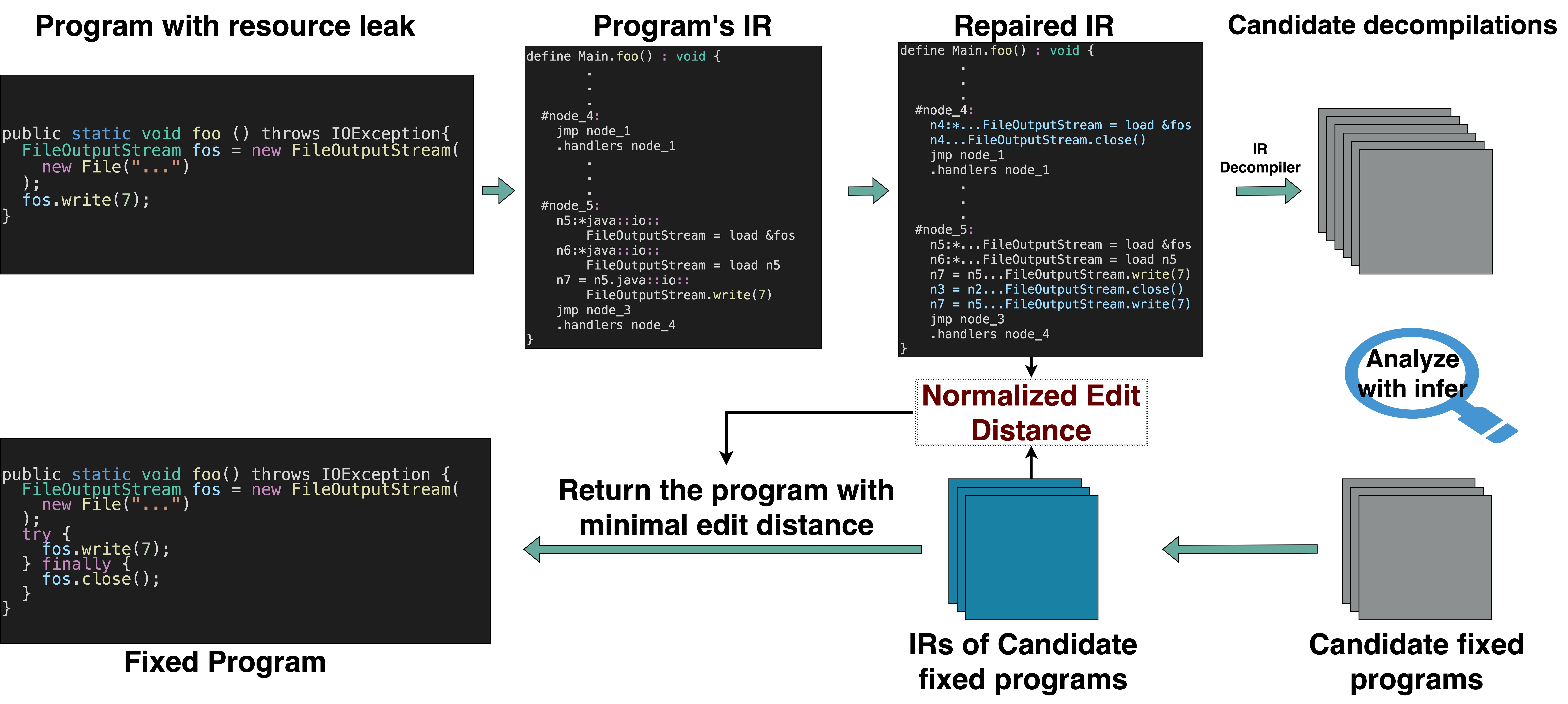}}
\caption{\small\textbf{Debugging using \infer IRs.} The code is compiled into the \infer IR, then fixed at the IR level and translated back to fixed source code using our model. We use the \infer and the java compiler to filter the generations sampled from the model. We use the normalized edit distance in IR space to select a single candidate to propose as a fix.}
\label{method_schema}

\end{center}
\vskip -0.2in
\end{figure}

\subsection{Data}
\label{sec:data}

\paragraph{\github BigQuery.} For our experiments, we use two data sets. First, there is the publicly accessible GitHub dataset on Google BigQuery, which consists of a corpus of GitHub files. Similarly to \citet{szafraniec2022code}, we only train on the portion of this dataset that can be compiled independently, so that we may create IRs to train the decompiler. This dataset was divided into three groups: training, validation, and testing.

\paragraph{\codenet.} \codenet~\citep{codenet} is a large-scale AI for code dataset for learning a variety of coding tasks. It has about 14 million samples of code, each of which is meant to be a solution to one of 4000 coding problems. We only use the java portion of the dataset, and we similarly filter out non-compiling files. For this dataset, we divided it into training, validation, and testing based on the problem ids. The java files in the dataset were submitted to 3219 problems, and we randomly selected 300 ids for validation, 300 for testing, and the remaining 2619 for training.

We use every Java file from our datasets for pre-training. 
For tasks involving Infer IRs such as IR decompilation, we limited ourselves to files that compile in isolation for practical reasons. 
The number of files in our dataset is shown in Table~\ref{tab:dataset_size}.

% \begin{table}[ht]
% \caption{\small\textbf{Dataset.}}
% \label{tab:dataset_size}
% \vskip 0.15in
% \begin{center}
% \begin{small}
% \begin{sc}
% \begin{tabular}{lccc}
% \toprule
% train & validation & test \\
% \midrule
% 94671 & 20278 & 24213\\
% \bottomrule
% \end{tabular}
% \end{sc}
% \end{small}
% \end{center}
% \vskip -0.1in
% \end{table}

\begin{table}[ht]
\caption{\small\textbf{Dataset.} This table shows the size of our dataset, in number of files, number of tokens and gigabytes. The full Java dataset is used for pre-training. 
The matched Java dataset corresponds to the subset of the full java dataset for which we were able to create an \infer IR. It is used for IR decompilation.
The \infer IR dataset contains the matching IRs. It is used for both pre-training and IR decompilation. }
\label{tab:dataset_size}
\vskip 0.15in
\begin{center}
\begin{small}
\begin{sc}
\begin{tabular}{lrrr}
\toprule
& Full Java & Matched Java & \infer IR\\
\midrule
NB files &	15.6~M &	1.6~M &	1.6~M \\
NB tokens &	16.5~B &	0.6~B & 6.4~B \\
Disk usage &	77.5~GB &	2.3~GB &	37.7~GB \\
\bottomrule
\end{tabular}
\end{sc}
\end{small}
\end{center}
\vskip -0.1in
\end{table}

\paragraph{Data preprocessing.} We use Tree-sitter\footnote{\url{https://tree-sitter.github.io/tree-sitter/}} to parse Java code into an AST and convert it to list of tokens. We tokenize \infer IRs by splitting them on white spaces and newlines. 
We used the same learned BPE~\citep{sennrich2015neural} codes as TransCoder~\citep{roziere2020unsupervised} and DOBF~\citep{roziere2021dobf}\footnote{\url{https://github.com/facebookresearch/CodeGen/tree/main/data/bpe/cpp-java-python}}. 
It is learned on AST-tokenized Java, C++ and Python code. 
We compile code using \texttt{javac}, get textual \infer representations using the \texttt{--dump-textual} options. %We use the options \texttt{--repair --pulse-only --debug-exceptions} when using \infer to find bugs and repair their IRs. 
% We learned BPE codes~\citep{sennrich2015neural} using \fastbpe\footnote{\url{https://github.com/glample/fastBPE}} on tokens % \fastbpe~\citep{} to tokenize the IRs after they have been generated, and AST tokenizer \citep{} is used first, then \fastbpe~\citep{} on the source code. Due to memory restrictions, we also discard samples that are longer than 2000 tokens.

\subsection{Model}
\label{sec:architecture}
\paragraph{Model architecture.} We consider a sequence-to-sequence (seq2seq) transformer~\citep{vaswani2017attention} model, which consists of an encoder and a decoder of 6 layers each and a total of about 312M parameters. We set the number of attention heads to 8, and the hidden dimension to 1024 (128 per head). We use layer dropout~\citep{fan2019layerdrop} with a probability parameter of 0.15 to make our training more efficient and reduce overfitting. 

\paragraph{Model pretraining.} Similarly to \citet{szafraniec2022code}, we pre-train our model using the denoising auto-encoding (DAE) task~\citep{vincent2008extracting} similarly to \citet{ahmad2021unified}, pre-trained using the masked language modeling task similarly to \citet{roziere2020unsupervised}. 
More precisely, we add noise to a sequence using a variety of procedures (e.g. shuffling, removal, masking subsequences) and train the model to retrieve the original sequence. 
We pre-train our model on our full Java dataset and on all the Infer IRs we were able to generate. 
DAE is an effective task to simultaneously train a model to understand IRs and generate Java source code. 
% The transformer architecture \citep{} that makes up our model of choice consists of an encoder and decoder networks with a total of about x parameters. We pre-trained the transformer model using the de-noising task, a supervised machine translation task in which a model is trained to predict a sequence of tokens given a corrupted version of that sequence. We use a variety of procedures (a combination of shuffling, removal, and random masking) to introduce noise into a sequence \citep{}, the DAE task honed the model’s "linguistic modeling" component. Additionally, we attempted to start the model from scratch and train it directly on the IR-inversion task, however this method resulted in a performance decrease and a quite slow rate of convergence.

% \subsection{IR inversion}
% \label{sec:method_ir_inversion}

\paragraph{IR inversion training.} For the IR inversion task, we generate parallel Java/IR sequences with \infer and train the model to reverse this process. 
More precisely, we feed our model with the \infer IR and train it to retrieve each subtoken of the corresponding Java source code using the cross-entropy loss.  
% A sequence-to-sequence methodology underlies the IR inversion task. Our model's input sequence x is an intermediate representation of the programme created by infer (Textual IR). 
% Given this input, The model is trained to retrieve the source code that produced the IR. The model parameters are learned by maximizing ground-truth likelihood during training time:
% At test time, for each stage of the decoding process, a distribution over the vocabulary is computed using the soft-max function. You may either use a greedy decoding strategy and pick the argmax at each step, or you can sample words from this distribution using some temperature and a sampling mechanism.

\paragraph{Code repair.} In order to perform code repair, the model is used to decompile the automatically corrected IRs of flawed programs into a bug-free version of these programs. The corrected IRs are obtained automatically using the method described in Section~\ref{sec:method_ir_fixes}. 
For this task, we favor improvements that require only minimal changes to the source code (buggy program). Therefore, for the decompilation of the repaired IRs, we select the best model using the edit distance.

% Tables 1, 2 detail the model's performance on the aforementioned measures as well as code repair.
\paragraph{Decoding strategies.} We evaluate our IR inversion model using greedy decoding. For the bug repair models, we use either beam search decoding~\citep{koehn2004pharaoh} or nucleus sampling~\citep{holtzman2019curiousnucleussampling} with temperature 1.3, p-value 0.9, and test several number of samples to get candidate programs from the decompiler. 
Before proposing a fix, we filter the generations to keep only those that compile with \texttt{javac} and do not raise a warning with \infer. If several samples pass this filter, we compute their IRs again and select the one with the minimal edit distance compared to the fixed IR. If no sample compiles without any \infer warning, we do not propose a fix. 
% In nucleus sampling, the emphasis is on the smallest set of the top words that together make up (100 * p-value)\% of the distribution. In doing so, you will account for the model's uncertainty about its prediction in the sampling process, meaning you will consider multiple tokens when the model is unsure and only stick to the top ones when the model is more confident. This is particularly helpful for code models where the compilation of a programme can fail with just one token.
% For the code repair task, we analyse the entire corpus of Java files using infer and keep the ones that were flagged with resource leaks; for these bugs, we generate the fixed IR and feed it to the decompiler; we sample candidate translations; we analyse the candidate/s with infer and detect whether the flag was suppressed for any of the translations; and we generate the IR again for the candidates that were successfully passing infer and compare it to the fixed IR to count the mean edit distance;  we also calculate the proportion that successfully compile with Javac out of the proposed candidates. 

\subsection{Automatic IR fixes}
\label{sec:method_ir_fixes}

Infer reports about a broad collections of programming errors such as deadlocks, buffer overflows, dubious dead code, null or dangling pointer dereferences, execution time regressions, etc\dots 
Among them, \textbf{resource leaks detection} catches subtle bugs and is amenable to program repair. This analysis tracks Java objects, the \textbf{resources}, that are supposed to be closed when not used any more in a program. Failure to close them is a \textbf{resource leak} that unnecessarily consume limited system resources (like file or socket handlers) and may lead to system crashes in production. Correct closing of such resources requires to add enough call to a \texttt{close} method, taking into account exceptional execution paths with \texttt{try/catch/finally} blocks.

For example, the following program is missing a call to \texttt{fos.close()} in the case where there is an exception during \texttt{fos.write()}. It will abruptly end the execution of \texttt{foo} without executing the close statement. 

\begin{minted}[fontsize=\small]{java}
public static void foo () throws IOException{
    FileOutputStream fos = new FileOutputStream(new File("..."));
    fos.write(7);
 }
\end{minted}

To fix this problem, the programmer can use a \texttt{finally} block that is executed in both normal and exceptional paths, as follow :
\begin{minted}[fontsize=\small]{java}
 public static void foo () throws IOException{
    FileOutputStream fos = new FileOutputStream(new File("..."));
    try {
      fos.write(7);
    } finally {
      fos.close();
    }
}
\end{minted}

At the IR level, resource leaks are detected at specific program points where all the following condition are met: i) the resource is still opened ;
ii) the resource is still reachable (in the heap) via a local variable \texttt{res};
iii) on all paths from the current point to an exit (normal or exceptional) of the current function, the 
variable \texttt{res} is not mentioned any more.
Such a program point is not only a right place to report for resource leak, but also to propose to add a close statement. Figure~\ref{fig:ir_to_fix_ir} in the Appendix presents an example of such a repair at IR level. The Java program at the top of the figure translates into the IR program on the left. Such a IR program is a list of instruction blocks. Each block has a label (\texttt{\#node\_0, ...}) and ends with jumps to other blocks (\texttt{jmp} instruction). The block \texttt{\#node\_1} is the exit block (with an empty list of successors). Some block also have a \textbf{handler} block (given with the keyword \texttt{.handlers}) which represents where the control flow will be rerouted if an exception is thrown. The program is missing two close statements and the IR on the right is the result of the automatic \infer fix. The first missing
close is added after the normal execution of the \texttt{write} call, in block \texttt{\#node\_5}. The second close is added in the block \texttt{\#node\_4} that is the handler block of block \texttt{\#node\_5}. If the \texttt{write} call ends with an exception, the rest of block  \texttt{\#node\_5} will not be executed and the control will jump to block \texttt{\#node\_4}. In such a situation, the method \texttt{foo()} will return  an exception but the resource is properly closed before that.

\subsection{Evaluation}
In this section, we define the metrics and methods used to evaluate our models for IR decompilation and bug repair. Except for the unit tests accuracy metric, we test the performance of our model for IR decompilation using the combined \codenet and \github BigQuery datasets. The unit test accuracy and the metrics for bug repair are evaluated using only the part of the \codenet for which we were able to extract working unit tests automatically.

\label{sec:evaluation_metrics}
\paragraph{IR inversion.}
We use the standard criteria for assessing machine translation models—perplexity, BLEU score~\citep{bleu}, and next token prediction accuracy—to evaluate the translation model's performance during training. Furthermore, we monitor a number of additional indicators to make sure the decompilation step introduces little to no modifications to the ground truth program's semantics. Concretely we evaluate the model on the following metrics, shown in Table~\ref{tab:IR_inversion_res}:

\begin{itemize}
    \item Mean normalized edit distance: the mean normalized edit distance (NED) between the ground truth program and the model's prediction. Normalized edit distance\textemdash as the name suggests\textemdash is the Levenshtein distance normalized by the length of the reference program.
    \item Compilation accuracy: the percentage of the model's output that compiles with Javac.
    \item IR exact match: percentage of retrieved source code that are compiles to exactly the same IR as the ground truth.
    \item Unit test accuracy: for the \codenet dataset, we parse the unit-tests from the problem description and only consider samples in the dataset that pass all their unit tests. We then decompile the IRs of those samples and run the unit tests on the output. The unit test accuracy is the percentage of the outputs that pass all the unit-tests. This metrics tests whether the generations semantically match the input. %This metric helps to get an idea of whether the model's generations have the same functional behavior as the ground truth.
\end{itemize}

\paragraph{Bug Repair.}
% To evaluate our model for bug repair, we identify 6540 files with resource leak alarm from the \codenet dataset and 370 from the BigQuery dataset. 
We evaluate our model for bug repair on CodeNet submissions with working unit tests.
We fix the resource leak automatically in IR space as described in Section~\ref{sec:method_ir_fixes}.
Then, we use our IR decompiler model to generate \texttt{N} candidate solutions from the fixed \infer IR, and run infer again to select successful fixes with no \infer alarm. 
Then, we select the top candidate among the successful fixes using the edit distance in IR space and run the unit tests.
We evaluate our model for bug repair using these metrics: 
\begin{itemize}
    \item Fix proposed: percentage of times when our model was able to propose at least one fixed version with no \infer alarm
    \item Fix precision: percentage of times when the proposed fix passes all the unit tests in CodeNet
    \item Fix recall: percentage of times when our system proposed a correct fix, i.e. there was at least a solution raising no alarm and the selected fix passes the unit tests. 
\end{itemize}

\section{Results and discussion}
\label{sec:results}

\begin{figure*}[t]\centering
    \begin{adjustbox}{width=1.2\textwidth,center}
\begin{tabular}{cc}
\toprule
  buggy program &  Proposed fix\\
\midrule
\begin{minipage}[t]{0.65\textwidth}
\begin{minted}{java}
void reader() throws IOException {
    FileInputStream fis = new FileInputStream("file.txt");
    BufferedInputStream bis = new BufferedInputStream(fis);
    try {
        bis.read();
        }
    finally {
    }
}

\end{minted}
\end{minipage}
&
\begin{minipage}[t]{0.65\textwidth}
\begin{minted}{java}
void reader() throws IOException {
    FileInputStream fis = new FileInputStream("file.txt");
    BufferedInputStream bis = new BufferedInputStream(fis);
    try {
        bis.read();
        }
    finally {
        bis.close();
    }
}
\end{minted}
\end{minipage} 
\\
% \\
% \begin{minipage}[t]{0.65\textwidth}
% \begin{minted}{java}
% public static void main (String [] args) {
%     Scanner stdIn = new Scanner(System.in);
%     BufferedInputStream bis = new BufferedInputStream(fis);
%     BufferedReader reader = new BufferedReader(
%         new InputStreamReader(System.in)
%         );
%     int a1 = stdIn.nextInt();
%     int a2 = stdIn.nextInt();
%     int a3 = stdIn.nextInt();
%     int min = a1;
%     int max = a1;
%     if (min > a2) min = a2;
%     if (min > a3) min = a2;
%     if (max < a2) max = a2;
%     if (max < a3) max = a2;
%     int ans = max - min;
%     System.out.println(ans);
% }
% \end{minted}
% \end{minipage}
% &
% \begin{minipage}[t]{0.65\textwidth}
% \begin{minted}{java}
% public static void main (String [] args) {
%     Scanner stdIn = new Scanner(System.in);
%     BufferedInputStream bis = new BufferedInputStream(fis);
%     BufferedReader reader = new BufferedReader(
%     new InputStreamReader(System.in));
%     reader.close()
%     int a1 = stdIn.nextInt();
%     int a2 = stdIn.nextInt();
%     int a3 = stdIn.nextInt();
%     int min = a1;
%     int max = a1;
%     if (min > a2) min = a2;
%     if (min > a3) min = a2;
%     if (max < a2) max = a2;
%     if (max < a3) max = a2;
%     int ans = max - min;
%     System.out.println(ans);
% }

% \end{minted}
% \end{minipage}
% \\
% \midrule

\bottomrule
\end{tabular}
\end{adjustbox}

\bigskip
\caption{\small \textbf{Examples of function fixed by our pipeline.} The function on the left has potential resource leaks. We use infer to create the corrected IRs for this function, and then we use the decompiler to translate the corrected IR back to the original source code. The outcomes is an edited version of the source code that has no resource leak. 
}
% TODO: can we find a better example for the first function? Here the reader is not used at all
\label{fig:example_fix}
\end{figure*}
 
Table~\ref{tab:IR_inversion_res} displays the results of the decompiler on the combined \codenet and \github BigQuery datasets. 
The version of our model pretrained with denoising auto-encoding reaches a slightly higher compilation rate than the model trained from scratch. 
It also reaches much higher scores on the IR match and unit test accuracy metrics, indicating that pretraining is especially beneficial for generating semantically valid decompiled code. 
Based on these metrics, we use the pretrained version of our decompiler for bug fixing. 

%For the unit test accuracy metric shown in the table, we extract unit tests from the 600 problem descriptions used in evaluations and testing and identify submissions that pass all of their unit tests in the validation and test portions of our datasets, only those submissions were used for the metric. 
%
% For the code repair task, we analyse the entire corpus of Java files using infer and keep the ones that were flagged with resource leaks; for these bugs, we generate the fixed IR and feed it to the decompiler; we sample candidate translations; we analyse the candidate/s with infer and detect whether the flag was suppressed for any of the translations; and we generate the IR again for the candidates that were successfully passing infer and compare it to the fixed IR to count the mean edit distance;  we also calculate the proportion that successfully compile with Javac out of the proposed candidates. 

% To determine the impact on the passing rate, we repeat this with various beam widths and numbers of random samples with nucleus sampling.

Table~\ref{tab:res_debugging_codenet} shows the performance of our model for fixing resource leak bugs from the \codenet dataset for nucleus sampling and beam search with several sample sizes. 
As expected, increasing the number of samples increases the number of proposed fixes for both decoding strategies. 
With our temperature value of 1.3 (selected through grid search), nucleus sampling starts much lower than beam search but its performance is essentially the same with sample size 100. 
The precision of the fix, measured using unit tests extracted from input and output examples from the \codenet problem statement, tends to decrease with the number of samples. 
However, the recall still increases with the number of samples due to the increase in number of proposed fixes. 
We hypothesize that, with more samples, our model starts proposing fixes for harder code snippets, therefore decreasing precision but increasing recall. 

Several methods could be studies to avoid the decrease in fix precision when increasing the number of samples. 
Our criterion to select the best candidate based on the edit distance between IRs may be lacking for large sample sizes. 
Additional methods, based on the number of the ratio of fix candidates that compile and raise no \infer warning could be used to decide whether to propose a fix. We could also develop better methods to evaluate the semantic equivalence of \infer IRs.
Overall, the decoding strategy and number of samples should depend on requirements in terms of latency, computational costs, precision and recall. 
While beam search seems clearly preferable for low sample sizes, sampling is likely to outperform it for sample sizes above 100. 
% It is clear from this table that when using lower number of samples, beam search yields better performance than nucleus sampling for the same number of samples, However this seizes to be the case as we grow the number of samples where in this case nucleus sampling produces similar or better results in terms of fix rate, which makes sense given how nucleus sampling affects prediction diversity compared to beam search with large beam sizes. With 200 samples, a temperature of 1.3, and a p-value of 0.9, we were able to effectively fix files out of a total of files with resource leak warnings, yielding a success rate of 68.8\%.

% the model is able to invert and find very interesting improvements.
Our decompiler also proposes interesting improvements to the coding style. 
Consider the second example of Figure~\ref{fig:decompilation_examples} in the Appendix: while the original code used the generic \texttt{Exception} class in the catch statement, the decompiled version of the example suggests using specific types of exceptions (i.e. \texttt{IOException} and \texttt{ClassNotFoundException}), which is generally regarded as better practice.  Similarly in the first example of Table~\ref{fig:decompilation_examples}, the model extracts the value \texttt{4096} into a constant \texttt{BUFFER\_SIZE = 4 * 1024}. 
In the third example, our model rewrites the modern try with resources\footnote{\url{https://docs.oracle.com/javase/tutorial/essential/exceptions/tryResourceClose.html}} construct into a more classical \texttt{try} syntax.
This type of refactors have no impact on the IR due to compilation-time optimizations.

\begin{table}[ht]
\caption{\small\textbf{IR inversion evaluations.}. This table shows our performance for decompiling \infer textual IR to Java source code. The token accuracy corresponds to the accuracy of the model when predicting the next token. 
NED stands for Normalized Edit Distance, or the Levenstein distance between the output and the ground truth divided by the length of the ground truth. These metrics were computed using greedy decoding. The unit test accuracy was computed on a subset of the \codenet dataset containing only the functions with working unit tests. 
The performance of the model trained from scratch is close to that of the pretrained model for the perplexity, compilation rate, BLEU score and edit distance, but much lower for the IR match and unit test accuracy metrics. Even though an IR match implies that the semantics of the function are the same, the unit test accuracy score is lower than the IR match score for the model from scratch because the unit tests are run only on a subset of the dataset. 
}
%\br{Do we know why the scores in test consistently better than in valid? }}
\label{tab:IR_inversion_res}
\vskip 0.15 in
\begin{center}
\begin{small}
\begin{sc}
\begin{tabular}{lcc}
\toprule
% Model & From scratch & Pretrained
Model & From scratch & Pretrained \\
\midrule 
Perplexity &  \scorethree{1.2211} &  \scorethree{1.1781} \\
BLEU &  \score{79.36} &  \score{78.85} \\
Token Acc &  \compacc{96.0721} &  \compacc{96.61} \\
Mean NED &  \scorethree{0.034498} &  \scorethree{0.03476} \\
Compil Rate &  \compacc{86.5333} &  \compacc{89.73} \\
IR match &  \compacc{60.6} &  \compacc{71.2} \\
Unit test Acc&  \compacc{59.4667} &  \compacc{86.9}  \\
\bottomrule
\end{tabular}
\end{sc}
\end{small}
\end{center}
\vskip -0.1in
\end{table}

\begin{table}[ht]
\caption{\small\textbf{Debugging results for \codenet submissions with working unit tests.} 
 We show the performance of our model for bug fixing with both nucleus sampling and beam search and for several sample sizes. As expected, the number of proposed fixes increases with the sample size, as it increases the probability to find a Java function which is not flagged by \infer. The fix precision decreases slightly with the number of samples and the recall increases. Nucleus sampling is performed with temperature \samplingtemp. These metrics are evaluated on 843 \codenet submissions with working unit tests. %\br{TODO Baptiste: get the value for beam search 200 or remove the line}
 \label{tab:res_debugging_codenet}}
\begin{center}
\begin{small}
\begin{tabular}{lccc}
\toprule
%Number of Translations & Sampling & Beam search & Unittest Acc\\
Decoding strategy & Fix proposed & Fix precision & Fix recall\\
\midrule
Sampling 1 & \compacc{56.82} & \compacc{97.29} & \compacc{55.28} \\
Sampling 5 & \compacc{64.53} & \compacc{97.43} & \compacc{62.87} \\
Sampling 10 & \compacc{66.31} & \compacc{97.32} & \compacc{64.53} \\
Sampling 50 & \compacc{67.97} & \compacc{97.03} & \compacc{65.95} \\
Sampling 100 & \compacc{68.33} & \compacc{96.70} & \compacc{66.07} \\
% Sampling 200 & \compacc{68.80} & \compacc{96.38} & \compacc{66.31} \\
\midrule
Beam Search 1 (Greedy) & \compacc{61.45} & \compacc{97.88} & \compacc{60.14} \\
Beam Search 5 & \compacc{65.60} & \compacc{97.65} & \compacc{64.06} \\
Beam Search 10 & \compacc{67.38} & \compacc{97.18} & \compacc{65.48} \\
Beam Search 50 & \compacc{68.56} & \compacc{96.89} & \compacc{66.43} \\
Beam Search 100 & \compacc{68.33} & \compacc{96.70} & \compacc{66.07} \\
% Beam Search 200 & - & - & - \\
\bottomrule
\end{tabular}
\end{small}
\end{center}
\end{table}

% \begin{table}[ht]
% \caption{\small\textbf{Debugging results for Codenets for 6540 res leak alarms}}
% \label{tab:res_debugging_codenet}
% \vskip 0.15in
% \begin{center}
% \begin{small}
% \begin{sc}
% \begin{tabular}{lccc}
% \toprule
% %Number of Translations & Sampling & Beam search & Unittest Acc\\
% N & Sampling & Beam search & Unittest Acc\\
% \midrule
% 1  & 3148 & 3288 & 69.18 \\
% 5  & 3623 & 3660 &  72.16  \\
% 10 & 3733 & - &  60.90  \\
% 50 & 3735 & 3986 &  57.49 \\
% 100 & 3983 & - &  53.47 \\
% % 200 & - & - &  51.47\\

% \bottomrule
% \end{tabular}
% \end{sc}
% \end{small}
% \end{center}
% \vskip -0.1in
% \end{table}

% \begin{table}[ht]
% \caption{\small\textbf{Debugging results for BigQuery for 370 res leak alarms}}
% \label{tab:res_debugging_bigquery}
% \vskip 0.15in
% \begin{center}
% \begin{small}
% \begin{sc}
% \begin{tabular}{lccc}
% \toprule
% Nb of Translations & Sampling & Beam search  \\
% \midrule
% 1  & 57 & 57  \\
% 5  & 71 & 71  \\
% 10 & 74 & 75  \\
% 50 & 83 & 80  \\
% 100 & 89 & 84  \\
% % 200 & 89\% & - \\
% \bottomrule
% \end{tabular}
% \end{sc}
% \end{small}
% \end{center}
% \vskip -0.1in
% \end{table}

\section{Related work}
\label{sec:related}

\subsection{Machine learning for automatic bug repair}
% Check this gdoc https://docs.google.com/document/d/1HZgvSMPfBkkD2SgMCYva4Rjic0qSO3Hli966NPdjmnY/edit?usp=sharing
\paragraph{Based on mined bug fixes.} 

Several studies use examples of manually fixed bugs to train machine learning algorithms. For instance,  \citep{tufano2019empirical} create a parallel dataset of buggy-fixed source code by selecting \github commits with words like "fix" or "solve" in their messages. Then, they train a model to translate from buggy to fixed code. SEQUENCER~\citep{chen2019sequencer} is a similar model, which specialized in one-line repairs. It is trained on a mined dataset of (buggy program, line number of the bug, fixed line). DeepFix \citep{gupta2017deepfix} trains a neural network to address compilation errors, they use a compiler as an oracle to validate patch candidates before suggesting them to the user.   

DeepRepair~\citep{white2019sorting} use auto-encoders to compute similarity scores between different code snippets to prioritize and transform repair components, and improve on redundancy-based program repair techniques. Another pattern-based method, called AVATAR~\citep{liu2019avatar}, uses fix patterns of static analysis violations as components for patch generation. Getafix~\citep{bader2019getafix} proposed a new hierarchical clustering technique that arranges fix patterns into a hierarchy ranging from broad to narrow patterns. This clustering approach is used to quickly choose the best fix to recommend to the user.

\citet{bhatia2018neuro} combine neural networks and constraint-based reasoning to repair syntax errors. \citet{dinella2020hoppity}~cast the problem of program repair as learning a sequence of graph transformations. Their model makes a sequence of predictions including the position of buggy nodes in the program's graph and corresponding graph edits to produce a fix for Javascript programs. \citet{vasic2019neural}~develop a multi-headed pointer networks for locating and correcting variable misuse bugs. DeepDebug \citet{drain2021deepdebug}~train a transformer model with back-translation, and fine-tune it using various program analysis information obtained from test suites.

\paragraph{Based on generated bugs.}
Other methods generate parallel buggy/fixed code pairs by introducing bugs automatically in correct code. For instance, BUGLAB~\citep{allamanis2021self} makes use of a selector model that decides which bug to apply to a code snippet and a detector model that detects and repair the inserted bug (if one was inserted). \citet{yasunaga2020graph} generate large amounts of bugs from correct code using several corruption rules. They train a graph-based neural network, dubbed Dr~Repair on the obtained data. 
\citet{ye2022selfapr} generate more than 1 million examples of training samples by perturbing code from open source projects. They outperform supervised repair approaches on Defects4J~\citep{just2014defects4j}.

\paragraph{Real-world datasets.} There are a few manually curated bug fix datasets for Java~\citep{just2014defects4j,durieux2016introclassjava,saha2018bugs,Madeiral2019bears,bui2022vul4j} and other languages~\citep{dbgbench,tan2017codeflaws,hu2019re}. Due to their small sizes, they are generally used only for evaluation purposes.

\paragraph{Large language models.}
Large language models trained on code have recently demonstrated strong capabilities for detecting and resolving bugs in code, with PALM~\citep{chowdhery2022palm} outperforming Dr~Repair on DeepFix~\cite{gupta2017deepfix}. They can also explain the bug and the proposed fix, which are useful features for programming assistants. \citep{prenner2021automatic,kolak2022patch} study the capabilities of such models for code debugging on quantitative benchmarks.

\subsection{Neural IR decompilation}
Other works used neural networks for decompilation. 
\citet{katz2019towards} train a sequence-to-sequence model to retrieve C code from either LLVM IRs or assembly code. 
\citet{fu2019coda} uses a LSTM model to generate code sketches for binary decompilation. In a second phase, these sketches are refined and corrected iteratively. 
\citet{liang2021semantics} trained a transformer model to decompile binaries to C. 
\citet{szafraniec2022code} leveraged LLVM IRs for source code translation, studying both IR inversion and IR-augmented pretraining methods. 
Similarly to our work, they consider IR decompilation to be a proxy objective for another task (code translation in their case). 

\subsection{Intermediate representations for static analysis}
Static analysis is often performed on intermediate representations instead on the source code itself. A first immediate advantage is that one single IR can serve several language front-ends. This is also a design choice that is popular in compilers (see for example the LLVM project~\citep{LLVM:CGO04}) which also perform static analysis, but for optimizations purposes rather than bugfinding. IR languages are much simpler than source languages: fewer syntactic constructs, and simpler language features. Analyzer designers save development
and maintenance time with such an architecture. 

\subsection{Automatic program repair}
Automatic program repair is a popular technique in IDE. For example, the Eclipse editor provides program repair for syntax errors. Our contribution deals with semantic error. For such errors, the dominant approach is based on testing~\citep{PerkinsKLABCPSSSWZER09}. When a program does not pass a test, a fixed program is searched and the success criterion is passing all the test suit. This approach can be directly driven at the source language level but it only observes a finite number of traces while static analysis can explore an infinite number by symbolic execution. A repair by static analysis is provided by the \texttt{ccheck} tool for C\# programs~\citep{LogozzoB12}. The static analysis is performed at the level of .NET bytecode language, which is a higher level representation than Infer IR, that deals with a strictly bigger set of frontend languages. The repaired properties they tackle are also of different nature because they require to modify a guard or adding an assignment in a linear block of instruction, while resource leak repair requires to build new control flow constructs like a \texttt{try {...} finally {...}} statement.

\section{Conclusion and future works}
\label{sec:conclusion}
Static analysis tools can flag potential issues in code, but they are often unable to propose suitable fixes. 
For instance, \infer produces an IR in which resource leaks can be found and fixed easily, but fixing the bug in the source code is difficult. 
We train a neural decompiler to retrieve the original source code from an \infer IR. 
This model generalizes to automatically fixed IRs, and it is able to propose proper fixes for up to 66.4\% of the code snippets in our test dataset. We measure the semantic correctness of our proposed fixes using unit tests and show that the precision varies between 96.4\% and 97.9\%. 
We believe that our method is precise enough to propose automatic PR for bug fixes, or automatic fixes in IDEs. 

% \paragraph{Future works.}
In this work, we only used \infer IRs built from Java files that compile in isolation. 
Integrating \infer to build tools and running it on entire projects would allow us to train the decompiler on much larger datasets and to improve its performance. 
The lack of correctness guarantees frequently hinders the adoption of machine learning models for code repair. 
Assuming that the compiler and our fixes in IR space are correct, it would be possible to get such guarantees for our system. 
The source code we produce can be compiled to IR again, and compared to the fixed IR. 
In our current setting, our fixed IR are unnatural and often do not match the IR produced by compiling any source code. 
Hence, we would like to develop automated methods to compare IRs semantically~\cite{Benton18} or even propose \textbf{semantic distances} between programs.
We also plan to extend this method to other types of bugs flagged by \infer, such as buffer overflows,  dead code, and null or dangling pointer dereferences.
% Another limitation is that, while our automatic fixes in IR space are semantically correct, they often do not match the IR produced by compiling any source code. 
% We would like to develop automated methods to compare IRs semantically. Then, by compiling our fixes again and comparing the new IR to the fixed IR, we would be able to guarantee that some of our fixes are correct. 

% Mention that we can improve the checking phase using methods to evaluate whether two IRs are semantically identical. 


\bibliography{biblio}
\bibliographystyle{plainnat}

%%%%%%%%%%%%%%%%%%%%%%%%%%%%%%%%%%%%%%%%%%%%%%%%%%%%%%%%%%%%%%%%%%%%%%%%%%%%%%%
%%%%%%%%%%%%%%%%%%%%%%%%%%%%%%%%%%%%%%%%%%%%%%%%%%%%%%%%%%%%%%%%%%%%%%%%%%%%%%%
% APPENDIX
%%%%%%%%%%%%%%%%%%%%%%%%%%%%%%%%%%%%%%%%%%%%%%%%%%%%%%%%%%%%%%%%%%%%%%%%%%%%%%%
%%%%%%%%%%%%%%%%%%%%%%%%%%%%%%%%%%%%%%%%%%%%%%%%%%%%%%%%%%%%%%%%%%%%%%%%%%%%%%%
\newpage
\appendix
\onecolumn
\section{Generation examples.}

% You can have as much text here as you want. The main body must be at most $8$ pages long.
% For the final version, one more page can be added.
% If you want, you can use an appendix like this one, even using the one-column format.
%%%%%%%%%%%%%%%%%%%%%%%%%%%%%%%%%%%%%%%%%%%%%%%%%%%%%%%%%%%%%%%%%%%%%%%%%%%%%%%
%%%%%%%%%%%%%%%%%%%%%%%%%%%%%%%%%%%%%%%%%%%%%%%%%%%%%%%%%%%%%%%%%%%%%%%%%%%%%%%

\begin{figure*}[h]\centering
    \begin{adjustbox}{width=1.0\textwidth,center}
\begin{tabular}{cc}
\toprule
  Original program &  Decompiled program\\
\midrule
\begin{minipage}[t]{0.65\textwidth}
\begin{minted}{java}
final class UploadRequest {
    private static void writeFileToOutputStream(
                                    File file, 
                                    OutputStream outputStream) 
                                    throws IOException {
        FileInputStream inputStream = null;
        try {
          inputStream = new FileInputStream(file);
          byte[] buffer = new byte[4096];
          int bytesRead;
    
          while ((bytesRead = inputStream.read(buffer)) > 0) {
            outputStream.write(buffer, 0, bytesRead);
          }
          outputStream.flush();
        
        } finally {
          if (inputStream != null) {
            inputStream.close();
          }
        }
      }
}
\end{minted}
\end{minipage}
&
\begin{minipage}[t]{0.65\textwidth}
\begin{minted}{java}
final class UploadRequest {
    private static final int BUFFER_SIZE = 4 * 1024 ;
    private static void writeFileToOutputStream(
                                    File file, 
                                    OutputStream outputStream) 
                                    throws IOException {
        FileInputStream inputStream = null;
        try {
          inputStream = new FileInputStream(file);
          byte[] buffer = new byte[BUFFER_SIZE];
          int bytesRead;
          while ((bytesRead = inputStream.read(buffer)) > 0) {
            outputStream.write(buffer, 0, bytesRead);
          }
          outputStream.flush();
          
        } finally {
          if (inputStream != null) {
            inputStream.close();
          }
        }
      }
}

\end{minted}
\end{minipage}
\\
\\
% \midrule
\begin{minipage}[t]{0.7\textwidth}
\begin{minted}{java}
static ArrayList<String> readListFromFile(File targetFile) {
    ArrayList<String> fileContents;
    try (ObjectInputStream inputStream = new ObjectInputStream(
                            new FileInputStream(targetFile))) {
      fileContents=(ArrayList<String>) inputStream.readObject();
      return fileContents;
    } catch (Exception e) {
      return new ArrayList<>();
    }
  }

\end{minted}
\end{minipage}
&
\begin{minipage}[t]{0.65\textwidth}
\begin{minted}{java}

static ArrayList<String> readListFromFile(File targetFile) {
    try {
        ObjectInputStream inputStream = new ObjectInputStream(
                                new FileInputStream(targetFile));
        ArrayList<String> 
        fileContents=(ArrayList<String>) inputStream.readObject();
        return fileContents;
    }
    catch (IOException e) {
        return new ArrayList<String>();
    }
    catch (ClassNotFoundException e) {
        return new ArrayList<String>();
    }
}

\end{minted}
\end{minipage} 
\\
\\
% \midrule
\begin{minipage}[t]{0.65\textwidth}
\begin{minted}{java}
private static void addFileToZip(
                        File base, 
                        String filePath, 
                        ZipOutputStream outZipStream) 
                        throws IOException {
    try (FileInputStream inputStream = new FileInputStream(
                            new File(base, filePath)); ) {
        byte[] buffer = new byte[BUFFER_SIZE_BYTES];
        outZipStream.putNextEntry(new ZipEntry(filePath));
        int bytes_read;
        while ((bytes_read = inputStream.read(buffer)) > 0) {
            outZipStream.write(buffer, 0, bytes_read);
        }
    } finally {
        outZipStream.closeEntry();
    }
}

\end{minted}
\end{minipage}
&
\begin{minipage}[t]{0.65\textwidth}
\begin{minted}{java}

private static void addFileToZip(
                        File base, 
                        String filePath, 
                        ZipOutputStream outZipStream) 
                        throws IOException {
    try {
        FileInputStream inputStream = new FileInputStream(
                                    new File(base, filePath));
        try {
            byte[] buffer = new byte[BUFFER_SIZE_BYTES];
            outZipStream.putNextEntry(new ZipEntry(filePath));
            int bytes_read;
            while ((bytes_read = inputStream.read(buffer)) > 0) {
                outZipStream.write(buffer, 0, bytes_read);
            }
        }
        finally {
            outZipStream.closeEntry();
        }
    }
    finally {
    }
}
\end{minted}
\end{minipage} 
\\
\bottomrule
\end{tabular}
\end{adjustbox}

\bigskip
\caption{\small \textbf{Example decompilations}  In this table, we highlight some examples of decompilations proposed by the decompiler. In this case, the input to the model is the IR of the original program in the left side of the table (without the repair step), and the output in the right side is the output of the model using greedy decoding. In these examples, we observe some style improvements in exception handling and variable assignment. The third example is a non-trivial decomposition of modern try-with-resources into multiple try finally statements.}
\label{fig:decompilation_examples}
\end{figure*}

\begin{figure}[h]
    \centering
    \begin{center}
\begin{minipage}[t]{0.8\textwidth}
\begin{minted}{java}
// Original function
public static void foo () throws IOException{
    FileOutputStream fos = new FileOutputStream(new File("/tmp/bar.txt"));
    fos.write(7);
}
    \end{minted}
    \end{minipage} 
    \end{center}
    \begin{adjustbox}{width=1.1\textwidth, center}
\begin{tabular}{cc}
\midrule
\begin{minipage}[t]{0.6\textwidth}
\scriptsize
\begin{minted}[escapeinside=||]{text}
# Infer IR
define Main.foo() : void {
  #node_0:
      jmp node_2
      .handlers node_1

  #node_1:
      jmp

  #node_4:
      jmp node_1
      .handlers node_1




  #node_2:
      n0 = __sil_allocate(<java::io::File>)
      n1 = java::io::File.<init>(n0, "/tmp/bar.txt")
      store &$irvar0 <- n0:*java::io::File
      jmp node_6
      .handlers node_4

  #node_6:
      n2 = __sil_allocate(<java::io::FileOutputStream>)
      n3:*java::io::File = load &$irvar0
      n4 = java::io::FileOutputStream.<init>(n2, n3)
      store &fos <- n2:*java::io::FileOutputStream
      jmp node_5
      .handlers node_4


  #node_5:
      n5:*java::io::FileOutputStream = load &fos
      n6:*java::io::FileOutputStream = load n5
      n7 = n5.java::io::FileOutputStream.write(7)
      jmp node_3
      .handlers node_4

  

  
  #node_3:
      jmp node_1
      .handlers node_4
}

\end{minted} 
  \end{minipage} & 
  \begin{minipage}[t]{0.55\textwidth}
  \scriptsize
\begin{minted}[escapeinside=||]{text}
# Fixed Infer IR
define Main.foo() : void {
  #node_0:
      jmp node_2
      .handlers node_1

  #node_1:
      jmp

  #node_4:
      |\colorbox{pastel_blue}{n4:*java::io::FileOutputStream = load &fos}|
      |\colorbox{pastel_blue}{n5 = n4.java::io::FileOutputStream.close()}|
      jmp node_1
      .handlers node_1

  #node_2:
      n0 = __sil_allocate(<java::io::File>)
      n1 = java::io::File.<init>(n0, "/tmp/bar.txt")
      store &$irvar0 <- n0:*java::io::File
      jmp node_6
      .handlers node_4

  #node_6:
      n2 = __sil_allocate(<java::io::FileOutputStream>)
      n3:*java::io::File = load &$irvar0
      n4 = java::io::FileOutputStream.<init>(n2, n3)
      store &fos <- n2:*java::io::FileOutputStream
      jmp node_5
      .handlers node_4

  #node_5:
      n6:*java::io::FileOutputStream = load &fos
      n7:*java::io::FileOutputStream = load n6
      n8 = java::io::FileOutputStream.write(n6, 7)
      |\colorbox{pastel_blue}{n2:*java::io::FileOutputStream = load &fos}|
      |\colorbox{pastel_blue}{n3 = n2.java::io::FileOutputStream.close()}|
      jmp node_3
      .handlers node_4

  #node_3:
      jmp node_1
      .handlers node_4

}
\end{minted} 
\end{minipage} \\
\midrule
\end{tabular}
\end{adjustbox}
    \centering
    \begin{center}
\begin{minipage}[t]{0.8\textwidth}
\begin{minted}{java}
// Fixed function
public static void foo () throws IOException{
    FileOutputStream fos = new FileOutputStream(new File("..."));
    try {
      fos.write(7);
    } finally {
      fos.close();
    }
}
    \end{minted}
    \end{minipage} 
    \end{center}
     \caption{\small \textbf{Resource leak fix in IR space.} For the original function is shown on top of the figure, we show the \infer IR on the left, and the automatically fixed \infer IR on the right. The lines added in the fixed IR are \colorbox{pastel_blue}{higlighted in blue}. 
     Using our decompiler on the \infer IR on the right, we obtain the fixed function on the bottom, which uses \texttt{finally} to safely close the resource. 
% DONE above (I have also removed the skip instruction because it does not play a role here)
%\br{TODO David: Could you add some explanations on why these fixes are needed and fix the resource leak? Is the \_\_sil\_skip() thing needed? More explanations could be useful. It's not obvious to me why we need to close the resource at the beginning of \#node4 but not in \#node2 for instance and I guess some readers will also have a hard time understanding.}
    }
    \label{fig:ir_to_fix_ir}

\end{figure}

\end{document}